\documentclass{PoS}

\usepackage{amsmath,amsthm, amssymb, latexsym}
\usepackage{dcolumn}
\usepackage{multirow}
\newcolumntype{d}{D{.}{.}{4.12}}
\newcommand*{\red}{\textcolor{red}}

\title{Kaon and D meson semileptonic form factors from lattice QCD}

\ShortTitle{Kaon and D meson semileptonic form factors from lattice QCD}

\author{\speaker{Thomas Primer}, Doug Toussaint\\
  Physics Department, University of Arizona, Tucson, AZ 85721, USA\\
        E-mail: \email{thomas.primer@gmail.com}}

\author{Aida El-Khadra\\
  Physics Department, University of Illinois, Urbana, IL 61801, USA}

\author{Elvira G\'{a}miz\\
  CAFPE and Departamenta de F\'{i}sica Te\'{o}rica y del Cosmos, Universidad de Granada, E-18071 Granada, Spain}

\author{James Simone\\
  Fermi National Accelerator Laboratory, Batavia, IL 60510, USA}

\author{Fermilab Lattice and MILC Collaborations}

\abstract{We present the status of on-going calculations of the $K\to\pi l\nu$ and $D\to K(\pi) l\nu$ semileptonic form factors at $q^2=0$.
These form factors are important for the determination of the CKM matrix elements $\lvert{V_{us}}\rvert$ and $\lvert{V_{cs(d)}}\rvert$ respectively.
This work uses the HISQ action for both valence quarks and sea quarks on MILC $N_f=2+1+1$ configurations.
We employ twisted boundary conditions to calculate the form factors at zero momentum transfer directly.
The $K\to\pi$ results are an update to previously published results with new data at the physical point.
The $D\to K(\pi)$ results are preliminary.}

\FullConference{The 32nd International Symposium on Lattice Field Theory\\
                 23-28 June, 2014\\
                 Columbia University New York, NY}

\begin{document}

\section{Introduction}

      The Standard Model predicts that the Cabbibo-Kobayashi-Maskawa (CKM) matrix,
      which describes the mixing between different flavors of quarks, must be unitary.
      Any deviation from unitarity would be an indication of new physics,
      and constraints can be established on the scale of new physics even if unitarity is fulfilled.

      The CKM matrix elements $\lvert{V_{us}}\rvert$, $\lvert{V_{cs}}\rvert$
      and $\lvert{V_{cd}}\rvert$ can be precisely determined
      from the semileptonic decays $K\to\pi l\nu$, $D\to K l \nu$ and $D\to\pi l \nu$ respectively.
      This determination can be done combining experimental results with lattice calculations of the corresponding vector form factors
      at zero momentum transfer, $f_+(0)$.

      Here we present a calculation of these form factors using HISQ valence quarks on MILC $N_f=2+1+1$ HISQ lattices.
      The details of the ensembles used are given in Table~\ref{ensembles}.

      \begin{table}[b]
        \centering
          \caption{Ensembles used in these calculations. \red{Red} indicates physical quark mass ensembles.}
          {
        \begin{tabular}{c c c c c c c c c}\hline\hline
          \multirow{2}{*}{$a$ (fm)} & \multirow{2}{*}{$m_l/m_s$} & \multirow{2}{*}{Volume} & \multicolumn{2}{c}{$N_{conf}\times N_{t_{source}}$}& 
           \multirow{2}{*}{$am_s^{sea}$} & \multirow{2}{*}{$am_s^{val}$} & \multirow{2}{*}{$am_c^{sea}$} & \multirow{2}{*}{$am_c^{val}$} \\
           & &  & ($K\to\pi$) & ($D\to K/\pi$) & & & & \\ \hline
          0.15 &\red{0.035}  & $32^3\times 48$ & $1000\times 4$ & & \red{0.0647} & \red{0.0691} & 0.831 & 0.8531 \\ \hline
          0.12    &  0.2     & $24^3\times 64$ & $1053\times 8$ & $1050 \times 8$ &0.0590 & 0.0535 & 0.635 & 0.6363 \\
                  &  0.1     & $32^3\times 64$ & $993 \times 4$ & $993 \times 4$ & 0.0507 & 0.053  & 0.628  & 0.650 \\
                  &  0.1     & $40^3\times 64$ & $391 \times 4$ & &0.0507 & 0.053  & 0.628  & 0.650 \\
            &  \red{0.035}   & $48^3\times 64$ & $945 \times 8$ & $943 \times 8 $ &\red{0.0507} & \red{0.0531} & 0.628 & 0.6269 \\ \hline
          0.09    &  0.2     & $32^3\times 96$ & $755 \times 4$ & $773 \times 4 $& 0.037  & 0.038 & 0.45 & 0.44 \\
                  &  0.1     & $48^3\times 96$ & $853 \times 4$ & $851 \times 4$ & 0.0363 & 0.038 & 0.44 & 0.43 \\
            &  \red{0.035}   & $64^3\times 96$ & $963 \times 8$ & $905 \times 8$ &\red{0.0363} & \red{0.0363} & 0.432 & 0.432 \\ \hline
          0.06    &  0.2     & $48^3\times 144$& $362 \times 4$ & &0.024  & 0.024 & 0.286 & 0.286  \\
            &  \red{0.035}   & $96^3\times 192$& $565 \times 6$ & $565 \times 6$ &\red{0.022}  & \red{0.022} & 0.26 & 0.26  \\ \hline \hline
          \end{tabular}\label{ensembles}}
        \end{table}

\section{Method}

      We do not calculate the vector form factor directly, but rather follow the method introduced in Ref.~\cite{Na:2010uf}
      which uses a Ward Identity to relate the matrix element of a vector current to that of a scalar current:
      \begin{equation}
        f_0^{DK}(q^2)=\frac{m_{c}-m_s}{m^2_{D}-m^2_K}\langle{K}\lvert{S}\rvert{D}\rangle_{q^2},
      \end{equation}
      together with the kinematic constraint $f_+(0)=f_0(0)$.
      The same applies to the other processes considered in these proceedings, $D\to\pi$ and $K\to\pi$.
      The main advantage of this approach over calculating the vector form factor directly is that the combination of
      $(m_{c}-m_s)S$ does not need renormalization.

\begin{figure}\centering
        \includegraphics[width=0.6\linewidth]{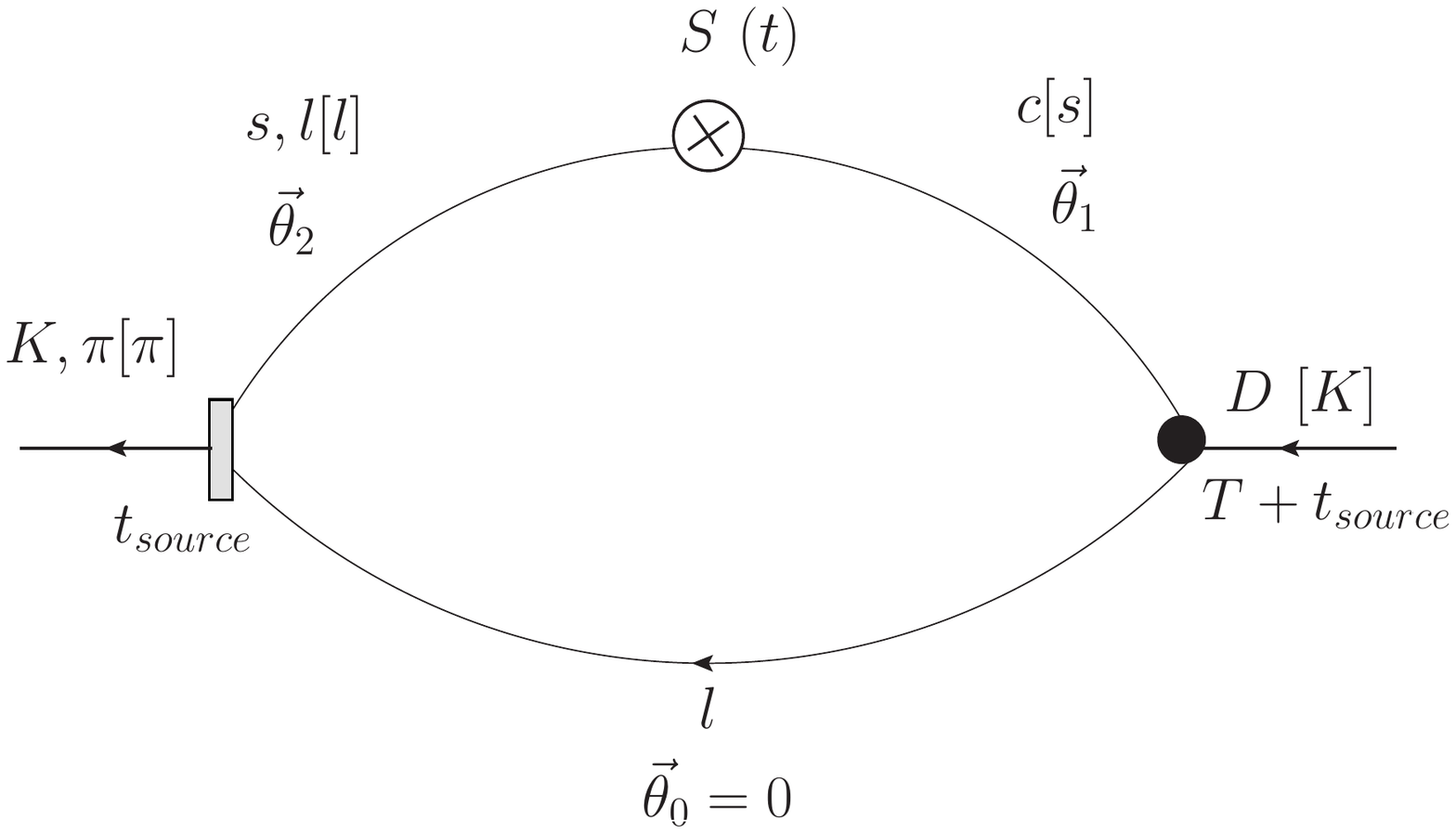}\vspace{-200pt}
        \caption{The structure of the 3pt correlators used for the calculation of the $D\to\pi$ and $D\to K$ scalar form factors,
          with the quantities in brackets corresponding to $K\to\pi$.}\label{3ptdiagram}
      \end{figure}

      Determination of the form factor requires the calculation of 2pt and 3pt correlation functions,
      where the 3pt correlators have the structure depicted in Figure~\ref{3ptdiagram}.
      We generate a light (strange) quark propagator at $t_{source}$ with random-wall sources and
      contract it with an extended strange (charm) propagator generated at $T+t_{source}$.

      In order to calculate 3pt correlators with zero momentum transfer we tune the momentum of the
      child particle using twisted boundary conditions.
      For each spatial direction $k$ with length $L$ these boundary conditions are defined as,
        \begin{equation}
          \psi(x_k + L) = e^{i\theta_k}\psi(x_k)
        \end{equation}
      This results in a propagator carrying momentum $p_k=\pi\frac{\theta_k}{L}$.
      Our initial $K\to\pi$ calculations experimented with putting the momentum on either the
      parent kaon or child pion ($\vec{\theta}_1\ne 0$ or $\vec{\theta}_2\ne 0$ in Fig~\ref{3ptdiagram} respectively),
      however the statistical errors were significantly larger with the momentum on the kaon.
      Therefore we now only put momentum on the child particle ($\vec{\theta}_1 = 0$ and $\vec{\theta}_2\ne 0$).

\section{$K\to\pi l \nu$ Updated Results}

\begin{figure}[b]
        \centering
        \includegraphics[width=0.6\linewidth]{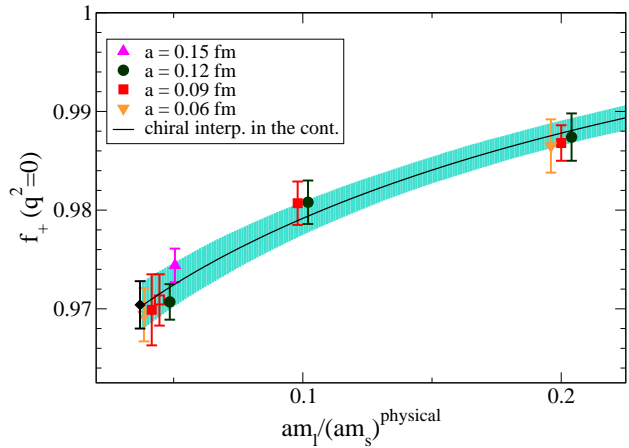}
        \caption{Chiral interpolation of $f_+^{K\pi}(q^2=0)$ from Ref.~\cite{Bazavov:2013maa}.
          The data points with open symbols show our preliminary new results
          for the $a\approx 0.06,\, 0.09$~fm physical mass ensembles.}
        \label{elviraextrap}
      \end{figure}

      Our results for $f_+^{K\pi}(0)$ and $\lvert{V_{us}}\rvert$ have been published previously in Ref.~\cite{Bazavov:2013maa}.
      In Fig.~\ref{elviraextrap} we show new preliminary results for the 0.06 fm physical quark mass ensemble
      as well as for the 0.09 fm physical mass ensemble with improved statistics in comparison
      with our previous results.
      The statistical error in the 0.09 fm point decreases from $~0.4\%$ to $~0.3\%$ when we double the number of time sources
      and add around 300 more configurations, while the central value is nearly unchanged.
      Our preliminary 0.06 fm result agrees very well with the other physical quark mass points,
      as well as with the extrapolation of Ref.~\cite{Bazavov:2013maa} value,
      and has similar errors.
      We are still generating data on this ensemble, so we expect to reduce the error shown in Fig.~\ref{elviraextrap}.

      Even when simulations at physical quark masses are available,
      application of a $\chi PT$ interpolation is useful for a number of reasons.
      Computationally cheaper data at $m_\pi > m_\pi^{phys}$ with high statistics can be included 
      to reduce the final statistical errors.
      Dominant discretization and finite volume effects can be analytically incorporated and removed.
      Small mass mistunings and partially quenched effects ($m_s^{sea}\ne m_s^{val}$) can be corrected at leading order.
      
      In order to arrive at a final result we use a combined chiral interpolation and continuum extrapolation.
      The form factor $f^{K\pi}_+(0)$ is written in terms of one-loop (NLO) partially quenched staggered $\chi PT$
      supplemented by two-loop (NNLO) continuum $\chi PT$ supplemented by analytic terms, 
      see Ref.~\cite{Bazavov:2013maa} for details.

      \begin{figure}\centering 
        \includegraphics[width=0.45\linewidth]{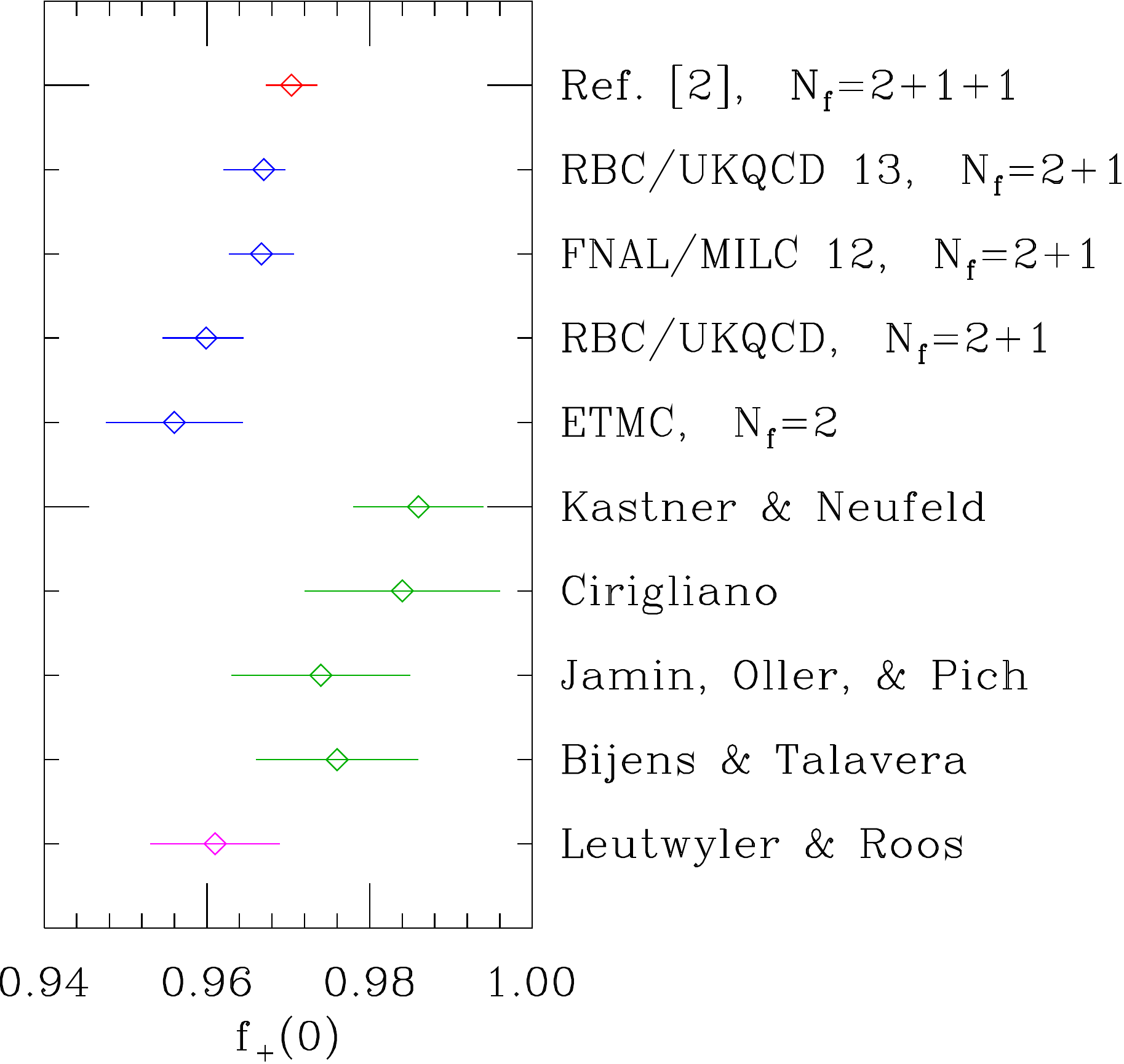}
        \includegraphics[width=0.3\linewidth]{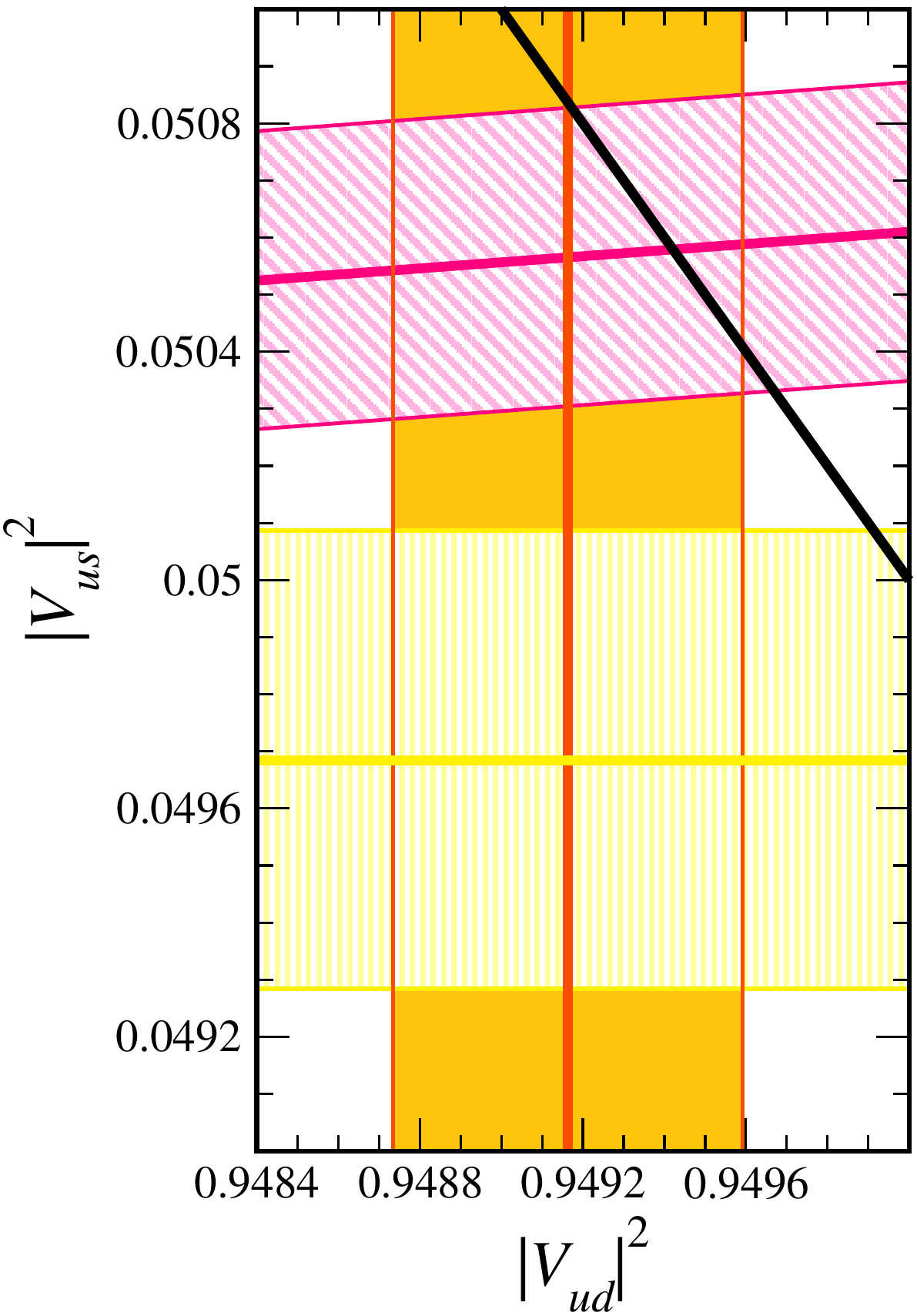}
        \caption{(left) Our value for $f_+^{K\pi}(0)$ from Ref.~\cite{Bazavov:2013maa} compared with previous work.
          (right) Bounds on first row unitarity using $V_{ud}$ from Towner \& Hardy,
          $f_k/f_\pi$ from Ref.~\cite{Bazavov:2014wgs} and $f^{K\pi}_+(0)$ from Ref.~\cite{Bazavov:2013maa}.} 
        \label{comparison}
      \end{figure}

      The result of the extrapolation for the form factor at physical quark mass in the continuum reported in Ref. \cite{Bazavov:2013maa} is:
      \begin{equation}
        f_+^{K\pi}(0)=0.9704(24)(22)=0.9704(32)
      \end{equation}
      Fig.~\ref{comparison} (left panel) shows a comparison of the result with previous lattice and non-lattice calculations.
      Combining it with the latest experimental average ($\lvert{V_{us}}\rvert f^{K\pi}_+(0)=0.2163(5)$ Ref.~\cite{Moulson:2013nsa})
      yields $\lvert{V_{us}}\rvert$ and subsequent check of first row unitarity:
      \begin{equation} \lvert{V_{us}}\rvert = 0.2290 \pm 0.00074_{lat} \pm 0.00052_{exp} \end{equation}
      \begin{equation}
        \rightarrow \Delta_{CKM} \equiv \lvert{V_{ud}}\rvert^2 + \lvert{V_{us}}\rvert^2 + \lvert{V_{ub}}\rvert^2 -1 =
        -0.00115(40)_{V_{us}}(43)_{V_{ud}} \label{bounds}
      \end{equation}
      where Eq.~\ref{bounds} uses the $\lvert{V_{ud}}\rvert$ determination of Ref.~\cite{Hardy:2013lga}.
      A graphical representation of the unitarity test is shown in the right panel of Fig.~\ref{comparison},
      which also includes the result of Ref.~\cite{Bazavov:2014wgs}.

\section{$D\to K(\pi) l \nu$ Current Status}

      The calculation of $D\to K$ and $D\to\pi$ follow the same process at that for $K\to\pi$.
      The main difference comes from the significantly larger mass of the $D$ meson compared to that of the kaon.
      This leads to a much larger momentum that must be injected into the kaon or pion to get $q^2=0$.
      As a result the moving kaon (pion) two-point correlators and $D\to K(\pi)$
      three-point correlators are quite noisy.
      The challenge then is in finding stable fits and keeping statistical and systematic errors under control.

      Our initial calculations focused on the physical quark mass ensembles at lattice spacings
      $a\approx 0.12$, $0.09$, and $0.06$ fm.
      We are currently expanding our analysis to a number of unphysical quark mass ensembles
      as indicated in Table~\ref{ensembles}.
      We can also analyse the form factor at $q^2_{max}$ (both particles stationary) without calculating any additional propagators.
      This allows us to investigate the dependence of statistical errors on the external momentum.
      
\begin{figure}\centering\vspace{-24pt}\hspace{-48pt}
        \includegraphics[width=0.50\linewidth]{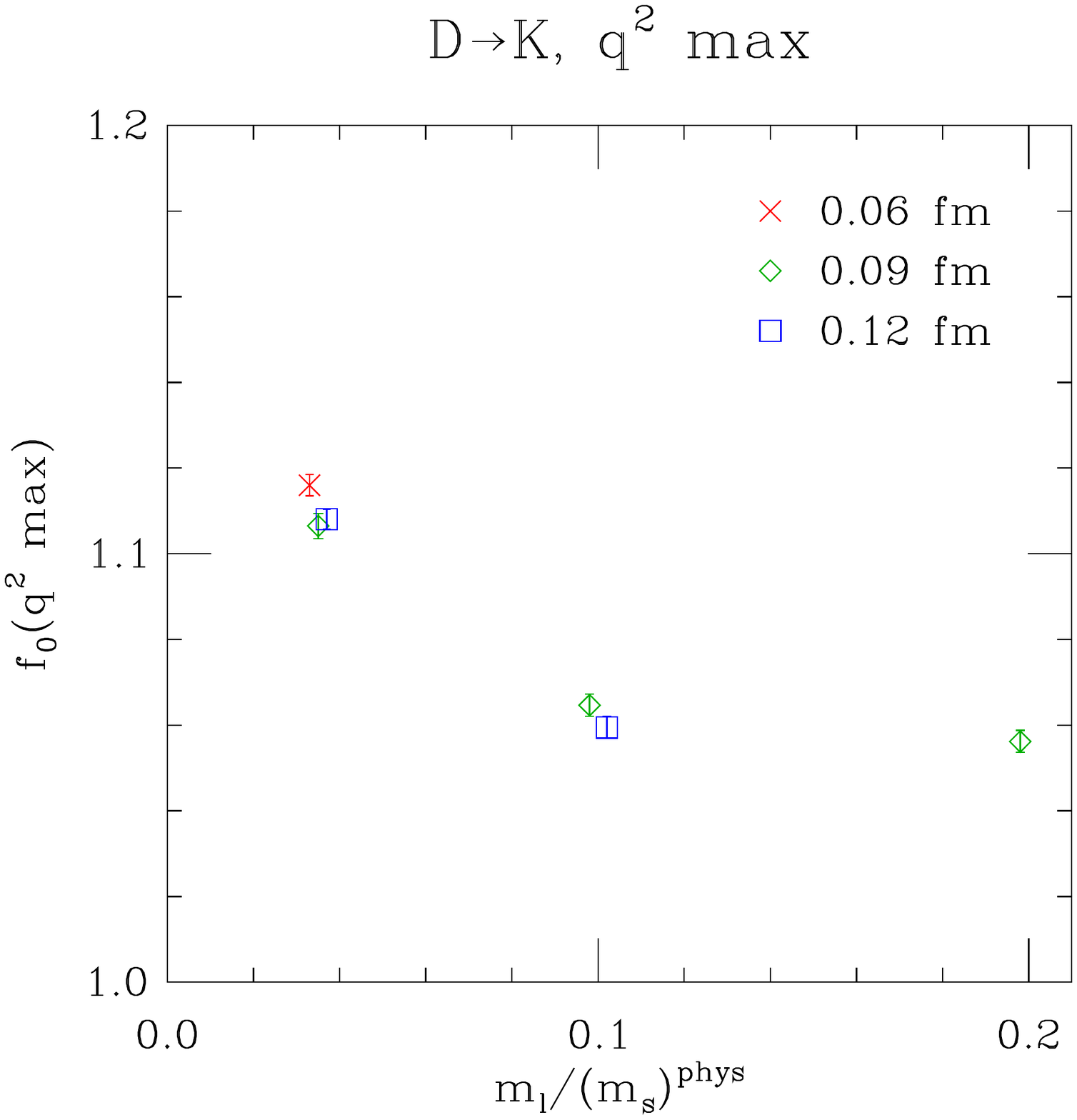}\hspace{-24pt}
        \includegraphics[width=0.50\linewidth]{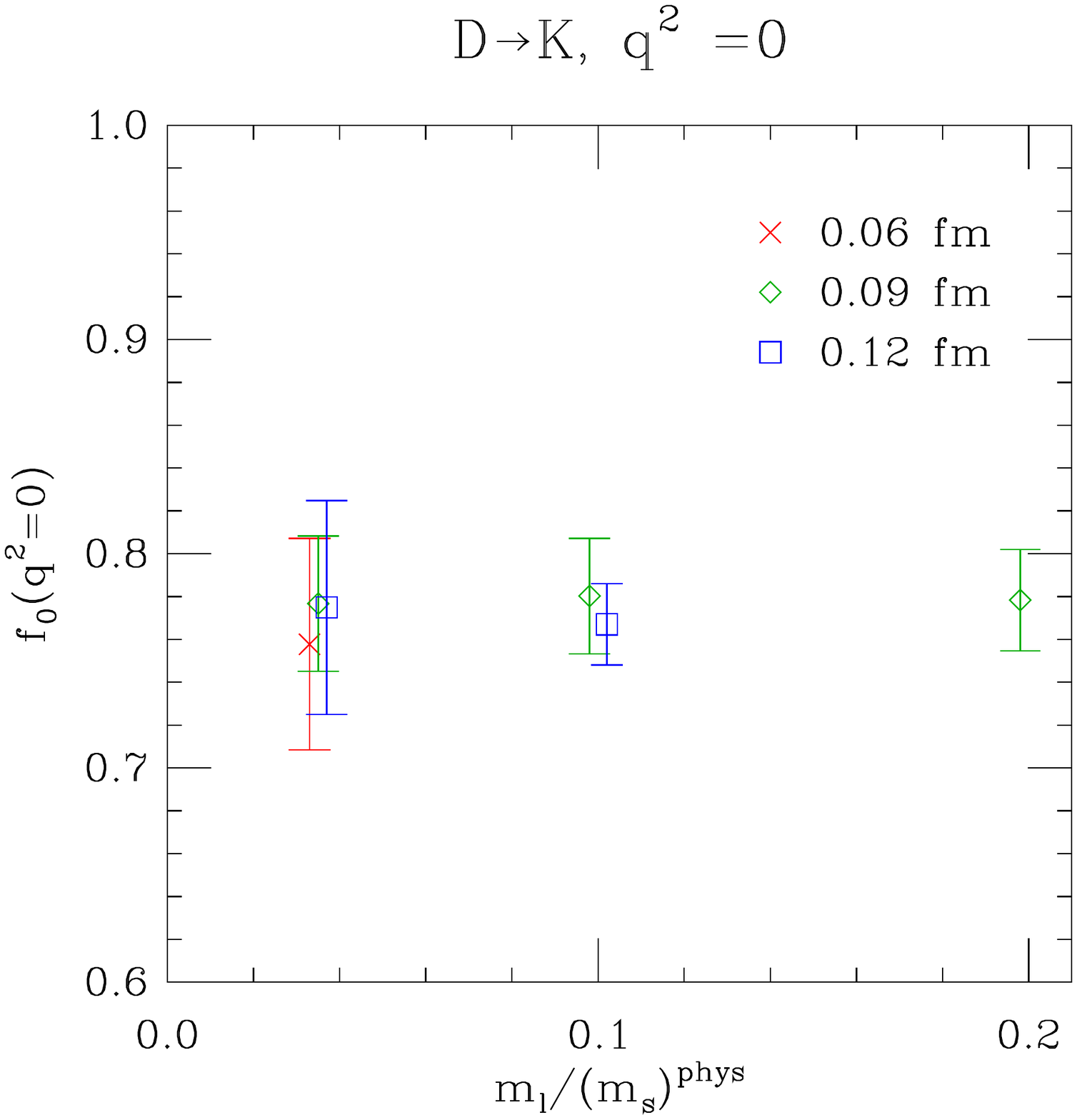}\\\vspace{-24pt}
        \caption{$f_0(q^2)$ for $D\to K$ as a function of quark mass at $q^2_{max}$ (left) and $q^2=0$ (right).}
        \label{dtok}
      \end{figure}

      The $D\to K$ form factor correlator fits are stable under variations of the fit parameters,
      which means that errors due to excited states and other fit systematics are under control.
      Our preliminary results for the form factor at $q^2=0$ and $q^2_{max}$ are shown in Fig.~\ref{dtok}
      as functions of light-quark mass.
      We observe that $f_0(q^2=0)$ shows no light-quark mass dependence within the statistical errors while $f_0(q^2_{max})$
      exhibits a statistically significant trend, increasing as the light-quark mass approaches its physical value.
      The analyses for other ensembles listed in Table~\ref{ensembles} are still in progress.
      The statistical errors for $f_0(q^2=0)$ are in general around 10 times as large as those for $f_0(q^2_{max})$.

      Finding good stability across fit windows and small statistical errors is difficult in the fits for $D\to\pi$ at $q^2=0$ 
      on a number of the ensembles.
      The large errors in the moving pion 2pt correlator and associated 3pt correlator leads to
      the usual $\chi^2$ per degree of freedom being a poor criterion for judging fits since almost all reasonable choices of fit window return
      a small $\chi^2/[dof]$ (or alternatively a $p$-value close to 1).
      These challenges will require more work to improve our fits.

\section{Future Plans}

      The preliminary results presented here for $f_+^{K\pi}(0)$ on two of our physical quark mass ensembles look promising
      for our goal of decreasing one of the main sources of uncertainty, statistical error, in our previous calculation.
      The reduction of the other dominant source of error, finite volume corrections, will require an explicit one-loop $\chi$PT calculation.

      Due to the large statistical errors that come with the momenta required to reach $q^2=0$ for D semileptonic decays,
      it is valuable to work with other values of $q^2$ as well.
      We plan to calculate $f_+^{DK(\pi)}(q^2)$ for a range of momentum values, which will require the evaluation of correlation functions
      with insertion of vectors currents, in addition to those with insertion of scalar currents.
      This will yield lattice form factors that cover the allowed range of $q^2$, which can be fit to a $z$-expansion \cite{Koponen:2013tua}
      to determine the shape.
      After checking the shape of the lattice form factor against experimental data we will use them to extract $\lvert{V_{cd(s)}}\rvert$
      in a combined fit.

\acknowledgments

This work was supported by the U.S. Department of Energy and National Science Foundation.
Computation for this work was done at the Argonne Leadership Computing Facility (ALCF),
the National Center for Atmospheric Research (UCAR),
Bluewaters at the National Center for Supercomputing Resources (NCSA),
the National Energy Resources Supercomputing Center (NERSC),
the National Institute for Computational Sciences (NICS),
the Texas Advanced Computing Center (TACC),
and the USQCD facilities at Fermilab, under grants from the NSF and DOE.
A.K.~is support by the U.S. Department of Energy under Grant No.~DE-FG02-13ER42001 and URA Visiting Scholars' program.
E.G.~is supported in part by MINECO (Spain) under Grants No.~FPA2010-16696 and No.~FPA2006-05294;
by Junta de Andaluc\'{\i}a (Spain) under Grants No.~FQM-101 and No.~FQM-6552;
and by the European Commission under Grant No.~PCIG10-GA-2011-303781.
Fermilab is operated by Fermi Research Alliance, LLC,
under Contract No.~DE-AC02-07CH11359 with the U.S. Department of Energy.

\end{document}